# Optical Emission Based Oxygen Sensing by ZnO Nanoparticles

Manoranjan Ghosh[*,†], R S Ningthoujam[#], R.K. Vatsa[#], D. Das[#], V. Nataraju[†], S.K. Gupta[†], D. Bahadur[‡]

*Bhabha Atomic Research Centre, Trombay, Mumbai 400085*
[*]Email: mghosh@barc.gov.in

ZnO is cheap, nontoxic and emerged as one of the most researched wide band gap (3.3 eV) optoelectronic materials due to its attractive optical properties.[1,2] Nanoparticles (NP) of ZnO is considered to be an useful material for sensing gases like $O_2$, $H_2S$, NO, $H_2$ etc.[3,4] Electrical resistance of aligned ZnO nanowires increases logarithmically as the oxygen gas pressure in the chamber is increased.[5] Generally, an optical technique is rarely used for the purpose of sensing gases. Vacuum system can be combined with an optical system by tracking the pressure dependent optical properties. For example refractive index of the gas inside a chamber is a function of a pressure therein.[6] Here we demonstrate that ZnO NP of size less than around 20 nm exhibit drastic fall in their visible photoluminescence (PL) intensity when oxygen pressure inside the sample chamber is decreased. The oxygen partial pressure dependent surface luminescence from ZnO NP demonstrated here has not been investigated earlier.

NP of ZnO show characteristic near band edge (NBE) emission in the UV region (~380nm) due to excitonic recombination and surface related visible PL in the wavelength range of 450–650nm[7]. The visible PL originates from defects such as charged oxygen vacancy, which are believed to be located near the surface.[8] Intensity of this broad emission is highly sensitive to the environment and mainly depends on the surface to volume ratio of the NP[12]. Hence, the sensitivity can be easily tuned by changing the particle size when this broad visible PL is employed to sense surrounding gases.

The phenomenon of oxygen pressure dependent visible PL is investigated on ZnO NP spread out as a film. For demonstration spherical NP of sizes 5, 10 and 15 nm have been chosen. These NP have been synthesized by dissociation of $Zn(CH_3COO)_2 \cdot 2H_2O$ in a basic medium (pH 11) at temperature 60-75 $^0$C.[9] A film (thickness 200-300nm) is obtained by adding NP dispersion in ethanol on a ITO (or quartz) substrate. An optical cryostat equipped with the provision of evacuation and passing select gases is used for PL measurement by a Spectrofluorimeter from Edinburg Instruments. For ZnO NP the wavelength range 320-355 nm is the most suitable window for excitation. For better visibility of NBE and visible bands, ZnO NP here is excited at 325 nm.

First we recorded the PL spectra of ZnO NP (dia~15nm) deposited on ITO or quartz substrate kept inside the optical cryostat. The sample chamber was initially at ambient air atmosphere. When evacuation starts the chamber pressure decreases, thereby reducing the visible PL intensity (figure 1). PL intensity in the range of 450-650 nm shows a sharp dependence within the tested range of chamber pressure. Almost complete reduction (95%) of visible PL is observed at highest vacuum level we could achieve $4.6 \times 10^{-5}$ mbar. Therefore, visible emission from ZnO NP can be very effectively controlled by changing the ambient air pressure. Hence, the pressure inside the chamber can be estimated by measuring PL intensity of ZnO NP from a known pressure dependent intensity curve.

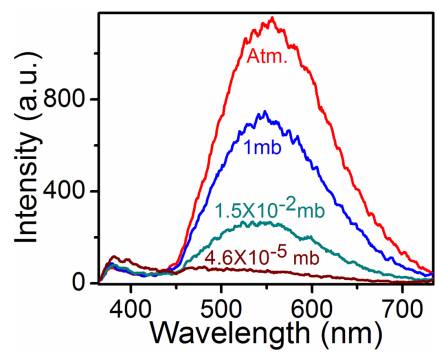

**Figure 1.** Photoluminescence spectra after excitation at 325 nm from a film of ZnO nanoparticles of size~15 nm for different air pressures.

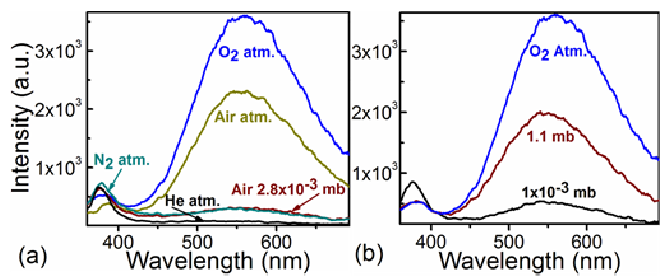

**Figure 2.** Photoluminescence spectra from ZnO nanoparticles of size~10 nm (a) in different environments and (b) for different oxygen pressures as indicated on the graph.

Since air is a mixture of gases, individual gases have been tested for their role in the phenomena described above. In figure 2(a), we show the PL spectra of ZnO NP (dia~10nm) kept in different gas environments. If the evacuated chamber is filled with oxygen, the PL intensity shoots up immediately (within 30 seconds) to 1.5 times of the value obtained in air atmosphere. This is because the chamber filled with oxygen has higher oxygen partial presure than that of air. Therefore oxygen partial pressure can be effectively sensed by the ZnO NP. However PL intensity of the evacuated chamber does not recover if it is filled by $N_2$ and He seperately. Therefore the visible PL intensity from ZnO are insensitive to the gases like $N_2$, He, Ar etc.

Once the decrease in partial pressure of oxygen is identified as the reason for reduction in emission intensity, we systematically collected PL spectra at various oxygen pressures. As shown in

---

[†]Technical Physics Division, BARC; [#]Chemistry Division, BARC
[‡] Indian Institute of Technology Bombay, Powai, Mumbai, 400076

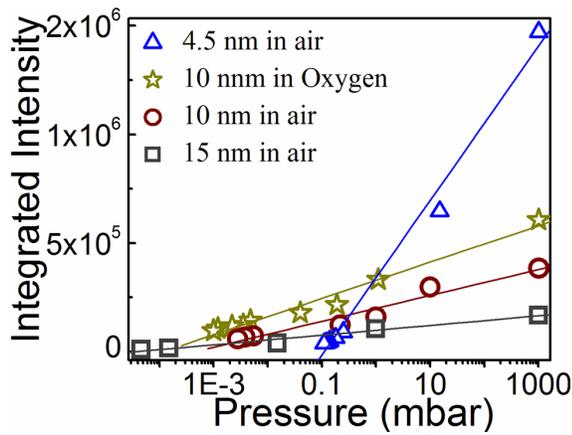

**Figure 3.** Integrated luminescence intensity shows approximately linear dependence with ambient pressure. Sensitivity increases when particle size decreases at the cost of sensing range of oxygen pressure.

figure 2(b), the intensity decreases as the oxygen pressure decreases as in the case of air. The intensity of the visible PL is proportional to the logarithm of the ambient chamber pressure (depicted by linear fitting of the data points in figure 3). Therefore the rate of change of PL intensity with pressure (or the sensitivity) is inversely proportional to the ambient air pressure. This establishes the fact that the investigated ZnO NPs are most suitable as low pressure sensor.

It would be interesting to compare the sensitivity or rate of change of intensity due to pressure for various situation discussed. From figure 3, it can be clearly seen that the sensitivity increases with the reduction in particle size. Therefore monitoring of the oxygen partial pressure by change of visible PL is a surface related phenomenon and depends on surface to volume ratio of the NP. The sensitivity (slope of linear fit) shows (1/diameter) dependence as expected. Further the sensitivity in pure oxygen environment is slightly higher than that in air. This is because the rate of decrease in partial pressure of oxygen in air during evacuation is slower than the situation when pure oxygen is considered. The range of oxygen partial pressure which can be sensed by this optical method depends clearly on the particle size. The range increases with the decrease in particle size at the cost of their sensitivity. The visible PL vanishes quickly for lower particle sizes. Therefore the sensitivity as well as the sensing range of oxygen partial pressure can be optimized as per the requirement by simply varying the particle size of the NP.

The sensing range of oxygen partial pressure also depends on the thickness of the film of ZnO NP up to a certain thickness. The visible PL vanishes within 8 mbar, 0.45 mbar and 0.31 mbar for NP film of thickness ~20nm, ~40nm and ~80nm respectively. There is no variation in the sensing range beyond the thickness of ~100 nm. The film thicknesses were kept within the range of 200-300 nm for all other samples investigated in this work. The sensitivity however does not show much variation with the thickness of the NP film.

In the line of the earlier studies we propose that the visible PL from ZnO NP originates from the charged oxygen vacancies which are predominantly located near the surface. Free-carrier depletion at the particle surface and its effect on the ionization state of the oxygen vacancy has strong impact on the visible emission[10]. An oxygen vacancy created in ZnO NP is neutral when not in contact with the surrounding fluid medium. The donor electrons in the conduction band of neutral oxygen deficient ZnO will occupy the acceptor like surface states created by excellent acceptor oxygen in atmosphere. By this way singly ($V_o^+$) and doubly ($V_o^{++}$) charged oxygen vacancies are created and the surface become positively charged. The existence of the positively charged depletion layer can be directly seen from the zeta potential (~18 mV) measured in these NP of small size dispersed in ethanol.[11] It can be shown that the broad visible emission in the blue yellow region is a composite of two lines located approximately at 2.2 eV (550 nm) and at 2.5 eV (500 nm).[12] Emission band appearing around 550 nm has been suggested to originate from $V_o^{++}$ whereas $V_o^+$ is responsible for the emission band around 500 nm.[13] The relative intensities and positions of these two lines determine the overall nature of the visible emission. The transfer of donor electrons to the acceptor like surface states forms a depletion layer near the surface. Such a depletion layer gives rise to band bending in nanospheres whose size is comparable to depletion width (~5–10 nm).[14] The extent of band bending in the near surface depletion layer in the NP is linked to their surface charge which decides the predominant nature of the visible emission. For positive charge on the surface, the band bending changes the chemical potential, which populates the level $Vo^{++}$ preferentially compared to the other level, leading to overall enhancement in the visible PL intensity. When oxygen is fully evacuated, no more electron transfer is possible. The ZnO NP having charged oxygen vacancies at the surface will be neutral ($Vo^X$). This modification of the surface charge by oxygen evacuation will change the band bending in such a way that there will be gradual reduction in the visible PL. Depending on the NP size, at sufficiently low oxygen pressure ZnO NP having neutral oxygen vacancy does not show surface related visible emission.

In conclusion, the phenomenon of reduction of the visible emission by lowering oxygen pressure opens up a way to exploit the potential of ZnO NP in low oxygen pressure measurement through an optical technique. The sensitivity can be improved by simply using smaller particles having higher intensity of the surface related emission band. The clear demonstration of charged oxygen vacancies to be responsible for the surface related visible emission from ZnO NP may resolve the long standing debate on this subject.

**Acknowledgement:** MG acknowledges the financial support from DST NANO Mission.

**Supporting Information Available:**. Synthesis of colloidal ZnO NP of different sizes and film of these NP on substrate. Characterization by TEM, XRD and AFM.

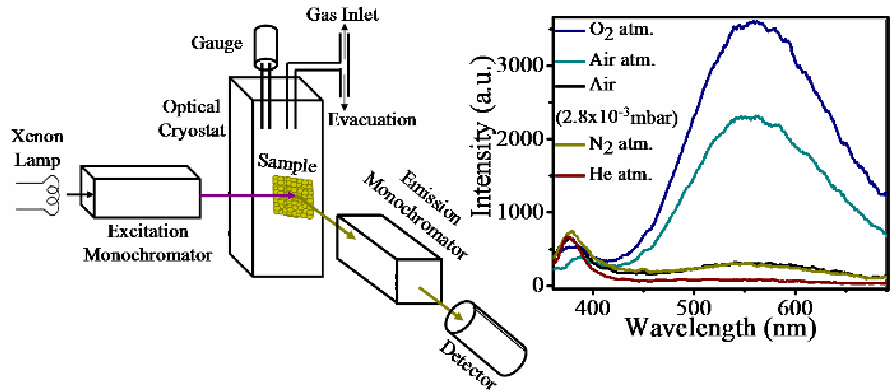

**Abstract:** ZnO nanoparticles (NP) of size less than around 20 nm inside a chamber exhibits complete reduction in their visible photoluminescence (PL) intensity when partial pressure of oxygen in the surrounding gaseous environment is decreased by evacuation. However the visible PL from ZnO nanoparticles are insensitive to the gases like $N_2$, He and Ar. The rate of change of PL intensity with pressure (or the sensitivity) is inversely proportional to the ambient air pressure. Sensitivity increases when surface to volume ratio increases for lower particle size. The charged oxygen vacancies are attributed as the responsible candidates for surface related visible PL from ZnO NP.